\documentclass[11pt,a4paper]{article}
\usepackage{amsmath}
\usepackage[mathcal]{eucal}
\usepackage{latexsym}
\usepackage{amssymb}
\begin{titlepage}
\title{Canonical quantization of Plebanski gravity in diagonal variables}
\author{Eyo Eyo Ita III}
\medskip
\input amssym.def
\input amssym.tex

\def \in{\indent}

\begin{document}
\maketitle
\bigskip
\centerline{Department of Applied Mathematics and Theoretical Physics} 
\smallskip
\centerline{Centre for Mathematical Sciences, University of Cambridge, Wilberforce Road}
\smallskip
\centerline{Cambridge CB3 0WA, United Kingdom}
\smallskip
\centerline{eei20@cam.ac.uk} 

\bigskip

\begin{abstract}
In this paper we have carried out a transformation from Ashtekar's theory of GR into a reduced theory where the physical degrees of freedom are explicit.  We have performed the canonical analysis, computed the classical dynamics and have performed a quantization on this reduced space, constructing a Hilbert space of states for vanishing cosmological constant.  Finally, we have clarified the canonical structure of the dual theory in relation to the original
Ashtekar theory.
\end{abstract}
\end{titlepage}

\section{Introduction}

The canonical formulation of the metric representation of general relativity produces a totally constrained system as a consequence of diffeomorphsim invariance.  The Hamiltonian consists of a linear combination of first class constraints $H_{\mu}=(H,H_i)$, respectively the Hamiltonian and diffeomorphism constraints.  These constraints $H_{\mu}$ have thus far turned out to be intractable in the metric representation due to their nonpolynomial structure in the basic variables.  A major development occured in 1988 with the introduction of the Ashtekar variables (see e.g. \cite{ASH1},\cite{ASH2},\cite{ASH3}), which led to the simplification of the initial value constraints into polynomial form.  The Ashtekar variables can be seen as a result of enlarging the metric phase space $\Omega$, essentially by embedding it into the phase space of a $SO(3)$ Yang--Mills theory.  A remnant of this embedding is the inclusion of the Gauss' law constraint $G_a$ in the list of constraints $H_{\mu}\rightarrow(H_{\mu},G_a)$.  The projection to the constraint shell has been problematic in the full theory also in the Ashtekar variables due to the presence of this additional constraint $G_a$\footnote{The spin network states of loop quantum gravity solve the Gauss' law constraint by construction, and provide a kinematic Hilbert $\boldsymbol{H}_{Kin}$ space for GR.  However, they have not yet to the author's knowledge been shown to solve the Hamiltonian constraint, which encodes the dynamics of the theory.  Still, many insights have resulted from the application of the Ashtekar variables at the classical and at the quantum level.}\par
\indent
In this paper we provide a prescription for projection from the full theory of the Ashtekar variables to the constraint shell through a series of transformations.  We then compute the Hamiltonian dynamics and carry out a quantization of the resulting reduced space.  The organization of this paper is as follows.  In section 2 we transform the Ashtekar action $I_{Ash}$ into a new action $I_{Inst}$ and then carry out the reduction in section 3 to the kinematic phase space $\Omega_{Kin}$.  We write the resulting action, which can be seen as $I_{Inst}$ at the level after implementation of the diffeomorphism and Gauss' law constraints.  In section 4 we formulate the canonical structure of the reduced action, transforming it into a canonical form exhibiting a cotangent bundle structure by restricting the configuration space $\Gamma_{Kin}$ to a diagonal connection.  It is found that the Hamiltonian constraint is a first class constraint, which enables the dynamics on $\Omega_{Kin}$ to be preserved.  In section 5 we compute the classical dynamics for $\Lambda=0$ and construct the spacetime metric, which is now a derived quantity.  In section 6 we carry out a quantization, constructing a Hilbert space of states annihilated by the Hamiltonian constraint for $\Lambda=0$.  This formalism enables the calculation of expectation values.  Section 7 establishes the canonical equivalence to the original Ashtekar variables, which highlights the role of the initial value constraints.  Section 8 is a brief summary and conclusion.

\section{Ashtekar variables into the instanton representation}

The action for general relativity in the Ashtekar variables can be written as the 3+1 decomposition of a totally constrained system, given by \cite{ASH1},\cite{ASH3}

\begin{eqnarray}
\label{ACTIONASH}
I_{Ash}=\int{dt}\int_{\Sigma}d^3x\widetilde{\sigma}^i_a\dot{A}^a_i+A^a_0D_i\widetilde{\sigma}^i_a\nonumber\\
-\epsilon_{ijk}N^i\widetilde{\sigma}^j_aB^k_a-{i \over 2}\underline{N}\epsilon_{ijk}\epsilon^{abc}\widetilde{\sigma}^i_a\widetilde{\sigma}^j_b\bigl(B^k_c+{\Lambda \over 3}\widetilde{\sigma}^k_c\bigr),
\end{eqnarray}

\noindent
where $\Lambda$ is the cosmological constant.  The basic phase space variables are a self-dual $SO(3,C)$ gauge connection $A^a_i$ and a densitized triad $\widetilde{\sigma}^i_a$.\footnote{The convention for labelling indices is that symbols from the beginning part of the Latin alphabet $a,b,c,\dots$ denote internal indices, while symbols from the middle $i,j,k,\dots$ denote spatial indices.}  The initial value constraints are $(G_a,H_i,H)$, the diffeomorphism, Hamiltonian and Gauss' law constraints, which are smeared by their respective Lagrange multiplier fields $(A^a_0,N^i,N)$.  These auxilliary fields are $A^a_0$, the temporal components of a four dimensional connection $A^a_{\mu}$, the shift vector $N^i$ and the lapse function $N$, and $\underline{N}=N(\hbox{det}\widetilde{\sigma})^{-1/2}$ is the densitized lapse function.\par
\indent
We will now perform a change of variables using the CDJ Ansatz \cite{CAP}

\begin{eqnarray}
\label{CEEANSATZ}
\widetilde{\sigma}^i_a=\Psi_{ae}B^i_e,
\end{eqnarray}

\noindent
where $B^i_a={1 \over 2}\epsilon^{ijk}F^a_{jk}$ is the magnetic field for $A^a_i$.  The matrix $\Psi_{ae}\in{SO}(3,C)\otimes{SO}(3,C)$, known as the CDJ matrix, is named after Riccardo Capovilla, John Dell and Ted Jacobson, 
and (\ref{CEEANSATZ}) is valid as long as $\Psi_{ae}$ and $B^i_a$ are nondegenerate three by three 
matrices.  Substitution of (\ref{CEEANSATZ}) into (\ref{ACTIONASH}) yields the action

\begin{eqnarray}
\label{ISGIVEN}
I_{Inst}=\int{dt}\int_{\Sigma}d^3x\Psi_{ae}B^i_e\dot{A}^a_i+A^a_0B^i_eD_i\Psi_{ae}\nonumber\\
-\epsilon_{ijk}N^iB^j_aB^k_e\Psi_{ae}-iN(\hbox{det}B)^{1/2}\sqrt{\hbox{det}\Psi}\bigl(\Lambda+\hbox{tr}\Psi^{-1}\bigr),
\end{eqnarray}

\noindent
which is defined on the phase space $\Omega_{Inst}=(\Psi_{ae},A^a_i)$.  To obtain (\ref{ISGIVEN}) we have used the Bianchi identity $D_iB^i_a=0$, combined with the characteristic equation for nondegenerate 3 by 3 matrices.\par
\indent
If (\ref{CEEANSATZ}) were a canonical transformation, then the phase space structure of (\ref{ISGIVEN}) would imply that the variable canonically conjugate to $\Psi_{ae}$ is an object $X^{ae}$ whose time 
derivative is $B^i_e\dot{A}^a_i$.  However, (\ref{CEEANSATZ}) is not a canonical transformation, which can be seen as follows.  The symplectic two form on the phase space $\Omega_{Ash}$ is given by

\begin{eqnarray}
\label{SYMPLEC}
\boldsymbol{\Omega}_{Ash}=\int_{\Sigma}d^3x{\delta\widetilde{\sigma}^i_a(x)}\wedge{\delta{A}^a_i(x)}=\delta\Bigl(\int_{\Sigma}d^3x\widetilde{\sigma}^i_a(x)\delta{A}^a_i(x)\Bigr)=\delta\boldsymbol{\theta}_{Ash},
\end{eqnarray}

\noindent
which is the exterior derivative of its canonical one form $\boldsymbol{\theta}_{Ash}$.  Using the functional Liebniz rule in conjuction with the variation 
of (\ref{CEEANSATZ}) we have $\delta\widetilde{\sigma}^i_a=B^i_e\delta\Psi_{ae}+\Psi_{ae}\delta{B}^i_e$, which transforms the left hand side of (\ref{SYMPLEC}) into

\begin{eqnarray}
\label{SYMPLEC1}
\boldsymbol{\Omega}_{Inst}=\int_{\Sigma}d^3x{\delta\Psi_{ae}}\wedge{B^i_e\delta{A}^a_i}+\int_{\Sigma}\epsilon_{ijk}\Psi_{ae}{\delta(D_jA^e_k)}\wedge{\delta{A}^a_i}.
\end{eqnarray}

\noindent
Due to the second term on the right hand side of (\ref{SYMPLEC1}), the symplectic two form for $I_{Inst}$ not in general exact and no such variable $X^{ae}$ exists on the phase space $\Omega_{Inst}$.  If there exist configurations where the second term of (\ref{SYMPLEC1}) vanishes, then such a canonical theory may be established.  We will obtain a canonical theory in two stages, starting with a reduction of (\ref{ISGIVEN}) to the kinematical phase 
space $\Omega_{Inst}\rightarrow\Omega_{Kin}$.  $\Omega_{Kin}$ is defined as the phase space after the diffeomorphism and the Gauss' law constraint have been implemented, leaving remaining the Hamiltonian constraint.

\section{Reduction to the kinematic phase space}

\noindent
The equation of motion for the shift vector $N^i$ implies that $\Psi_{ae}=\Psi_{ea}$ is symmetric.  Using the relation $F^a_{0i}=\dot{A}^a_i-D_iA^a_0$ for the temporal component of the curvature and performing an integration by parts in conjunction with the Bianchi identity, (\ref{ISGIVEN}) for symmetric $\Psi_{ae}$ reduces to

\begin{eqnarray}
\label{IMPLEMENT}
I_{Inst}=\int{dt}\int_{\Sigma}d^3x\Bigl[{1 \over 2}\Psi_{(ae)}\epsilon^{ijk}F^a_{0i}F^e_{jk}-iN(\hbox{det}B)^{1/2}\sqrt{\hbox{det}\Psi}\bigl(\Lambda+\hbox{tr}\Psi^{-1}\bigr)\Bigr].
\end{eqnarray}

\noindent
Equation (\ref{IMPLEMENT}) can be written in covariant form using the definition $\epsilon^{ijk}\equiv\epsilon^{0ijk}$, and invoking the symmetries of the 4-dimensional epsilon tensor $\epsilon^{\mu\nu\rho\sigma}$.  The resulting action is given by

\begin{eqnarray}
\label{IMPLEMENT4}
I_{Inst}=\int_Md^4x\Bigl[{1 \over 8}\Psi_{ae}F^a_{\mu\nu}F^e_{\rho\sigma}\epsilon^{\mu\nu\rho\sigma}-iN(\hbox{det}B)^{1/2}\sqrt{\hbox{det}\Psi}\bigl(\Lambda+\hbox{tr}\Psi^{-1}\bigr)\Bigr],
\end{eqnarray}

\noindent
where $F^a_{\mu\nu}=\partial_{\mu}A^a_{\nu}-\partial_{\nu}A^a_{\mu}+f^{abc}A^b_{\mu}A^c_{\nu}$ is the curvature of the four dimensional connection $A^a_{\mu}$.  Since $\Psi_{ae}$ is symmetric we can write it as a polar decomposition\footnote{We assume that $\Psi_{ae}$ is diagonalizable, which requires the existence of three linearly independent eigenvectors \cite{WEYL}.  Additionally, we will assume that the eigenvalues are nonzero.} 
 
\begin{eqnarray}
\label{INSTANT1}
\Psi_{ae}=(e^{\theta\cdot{T}})_{af}\lambda_f(e^{-\theta\cdot{T}})_{fe},
\end{eqnarray}

\noindent
using a $SO(3,C)$ transformation $(e^{\theta\cdot{T}})_{ae}$ parametrized by three complex angles $\vec{\theta}=(\theta^1,\theta^2,\theta^3)$.  This corresponds to a rotation of the 
diagonal matrix of eigenvalues $\lambda_f=(\lambda_1,\lambda_2,\lambda_3)$ from the intrinsic frame, where $\Psi_{ae}$ is diagonal, into an arbitrary $SO(3,C)$ frame.  
Substitution of (\ref{INSTANT1}) into the first term of (\ref{IMPLEMENT4}) yields

\begin{eqnarray}
\label{INSTANT2}
I_1={1 \over 8}\int_Md^4x\lambda_f((e^{-\theta\cdot{T}})_{fa}F^a_{\mu\nu}[A])((e^{-\theta\cdot{T}})_{fe}F^e_{\rho\sigma}[A])\epsilon^{\mu\nu\rho\sigma}.
\end{eqnarray}

\noindent
Note that the internal index on each curvature in (\ref{INSTANT2}) has been rotated by $e^{-\theta\cdot{T}}$, which corresponds to a $SO(3,C)$ gauge transformation.  Therefore there exists a 
curvature $f^a_{\mu\nu}[a]=(e^{-\theta\cdot{T}})_{ae}F^e_{\mu\nu}[A]$ corresponding to some four dimensional connection $a^a_{\mu}$.  The relation between $a^a_{\mu}$ and $f^a_{\mu\nu}$, which contains no explicit reference 
to the $SO(3,C)$ angles $\vec{\theta}$, is given by $f^a_{\mu\nu}=\partial_{\mu}a^a_{\nu}-\partial_{\nu}a^a_{\mu}+f^{abc}a^b_{\mu}a^c_{\nu}$.  It then follows that the connection $a^a_{\mu}$ is a $SO(3,C)$ gauge transformed version of $A^a_{\mu}$ related by

\begin{eqnarray}
\label{INSTANT5}
a^a_{\mu}=(e^{-\theta\cdot{T}})_{ae}A^e_{\mu}-{1 \over 2}\epsilon^{abc}(\partial_{\mu}(e^{-\theta\cdot{T}})_{bf})(e^{-\theta\cdot{T}})_{cf},
\end{eqnarray}

\noindent
which corresponds to the adjoint representation of the gauge group \cite{GAUGE}.  Defining $b^i_a={1 \over 2}\epsilon^{ijk}f^a_{jk}$ as the magnetic field of $a^a_i$, and using the complex orthogonal 
property $\hbox{det}(e^{\theta\cdot{T}})=1$, then (\ref{IMPLEMENT4}) can be written as

\begin{eqnarray}
\label{IMPLEMENT41}
I_{Inst}=\int_Md^4x\Bigl[{1 \over 8}\lambda_ff^f_{\mu\nu}f^f_{\rho\sigma}\epsilon^{\mu\nu\rho\sigma}
-iN(\hbox{det}b)^{1/2}\sqrt{\lambda_1\lambda_2\lambda_3}\Bigl(\Lambda+{1 \over {\lambda_1}}+{1 \over {\lambda_2}}+{1 \over {\lambda_3}}\Bigr)\Bigr]
\end{eqnarray}

\noindent
where we have used $B^i_a=(e^{\theta\cdot{T}})_{ae}b^i_e$ as well as the cyclic property of the trace.  The 3+1 decomposition of (\ref{IMPLEMENT41}) is given by

\begin{eqnarray}
\label{INSTANT6}
I_{Inst}=\int{dt}\int_{\Sigma}d^3x\Bigl[\lambda_fb^i_f\dot{a}^f_i+a^f_0b^i_f\underline{D}_i\{\lambda_f\}
-N(\hbox{det}b)^{1/2}\sqrt{\lambda_1\lambda_2\lambda_3}\Bigl(\Lambda+{1 \over {\lambda_1}}+{1 \over {\lambda_2}}+{1 \over {\lambda_3}}\Bigr)\Bigr],
\end{eqnarray}

\noindent
where $\underline{D}_i$ is the covariant derivative with respect to the connection $a^a_i$.  Variation of $a^f_0$ in (\ref{INSTANT6}) would result on an additional constraint on $\lambda_f$ which is unsatisfactory, since we would like to use $\lambda_f$ for the physical degrees of freedom of the theory.  To avoid this, we will now choose $a^f_0=0$, which also has the effect of eliminating three unphysical degrees of freedom.\par
\indent
The effect of the choice $a^f_0=$ will be to decouple the Gauss' law constraint from the reduced space $\Omega_{Kin}$.  However, the Gauss' law constraint can still be implemented on the larger phase space $\Omega_{Inst}$ by 
variation of $A^a_0$ in (\ref{ISGIVEN}).  Combined with the decomposition (\ref{INSTANT1}), this yields

\begin{eqnarray}
\label{GOOOSE}
G_a=B^i_eD_i\{(e^{\theta\cdot{T}})_{af}\lambda_f(e^{-\theta\cdot{T}})_{fe}\}=0
\end{eqnarray}

\noindent
which is a triple of differential equations.  For each $(\lambda_1,\lambda_2,\lambda_3)$ and $A^a_i$, (\ref{GOOOSE}) should in principle fix the angles $\vec{\theta}=\vec{\theta}[\vec{\lambda},A]$.  Note that the choice 
$A^a_0=\epsilon^{abc}(e^{-\theta\cdot{T}})_{bf}{d \over {dt}}(e^{-\theta\cdot{T}})_{cf}$ is consistent with $a^f_0=0$.  Hence one first implements the Gauss' law constraint on $\Omega_{Inst}$, following by projection to $\Omega_{Kin}$ by choosing $\vec{\theta}$ in the decomposition (\ref{INSTANT1}) to be the solution to (\ref{GOOOSE}).  Then the action on the kinematic phase space is given by

\begin{eqnarray}
\label{INSTANT7}
I_{Kin}=\int{dt}\int_{\Sigma}d^3x\Bigl[\lambda_fb^i_f\dot{a}^f_i
-iN(\hbox{det}b)^{1/2}\sqrt{\lambda_1\lambda_2\lambda_3}\Bigl(\Lambda+{1 \over {\lambda_1}}+{1 \over {\lambda_2}}+{1 \over {\lambda_3}}\Bigr)\Bigr].
\end{eqnarray}

\section{Canonical structure on the kinematic phase space}

\noindent
We will now compute the classical dynamics of the reduced theory on $\Omega_{Kin}$.  Appending a factor of $-{i \over G}$, the action is given by

\begin{eqnarray}
\label{INSTANT711}
I_{Kin}
=-{i \over G}\int{dt}\int_{\Sigma}d^3x\Bigl[\lambda_fb^i_f\dot{a}^f_i
-iN(\hbox{det}b)^{1/2}\sqrt{\lambda_1\lambda_2\lambda_3}\Bigl(\Lambda+{1 \over {\lambda_1}}+{1 \over {\lambda_2}}+{1 \over {\lambda_3}}\Bigr)\Bigr].
\end{eqnarray}

\noindent
Recall that the initial phase space $\Omega_{Inst}$ was of dimension $(9,9)$, namely with $9$ momentum and $9$ configuration space degrees of freedom per point.  Implementation of the diffeomorphism and the Gauss' law constraints respectively resulted in the following reduction sequence

\begin{eqnarray}
\label{REDUCTIONS}
Dim(\Omega_{Inst})=(9,9)\longrightarrow(6,9)\longrightarrow(3,6).
\end{eqnarray}

\noindent
The configuration space $a^a_i$ in (\ref{INSTANT711}) contains three more degrees of freedom per point than the momentum space $(\lambda_1,\lambda_2,\lambda_3)$.  To have a cotangent bundle structure on the reduced space we must eliminate three degrees of freedom from $a^f_i$.  Let us set three elements of $a^f_i$ to zero, by choosing a diagonal connection

\begin{displaymath}
a^a_i=
\left(\begin{array}{ccc}
a_1 & 0 & 0\\
0 & a_2 & 0\\
0 & 0 & a_3\\
\end{array}\right)
; b^i_e=
\left(\begin{array}{ccc}
a_2a_3 & -\partial_3a_2 & \partial_2a_3\\
\partial_3a_1 & a_3a_1 & -\partial_1a_3\\
-\partial_2a_1 & \partial_1a_2 & a_1a_2\\
\end{array}\right)
,
\end{displaymath}

\noindent
where $a_f=a_f(x,t)$ contain three independent degrees of freedom per point (and therefore corresponds to the full theory and not minisuperspace).  This particular configuration corresponds to a canonical one form

\begin{eqnarray}
\label{SYMPSTRUC}
\boldsymbol{\theta}=\int_{\Sigma}d^3x\Bigl(\lambda_1a_2a_3\delta{a}_1+\lambda_2a_3a_1\delta{a}_2+\lambda_3a_1a_2\delta{a}_3\Bigr).
\end{eqnarray}

\noindent
Note that there are no spatial gradients in (\ref{SYMPSTRUC}), which is a consequence of the fact that the spatial gradients in $b^i_a$ are contained in the off-diagonal positions of the matrix.  The variation of (\ref{SYMPSTRUC}) yields

\begin{eqnarray}
\label{SYMMPL}
\delta\boldsymbol{\theta}=\int_{\Sigma}d^3xa_2a_3{\delta\lambda_1}\wedge{\delta{a}_1}+\lambda_1{\delta(a_2a_3)}\wedge{\delta{a}_1}+Cyclic~Perms.
\end{eqnarray}

\noindent
which does not yield a symplectic two form of canonical form.  To remedy this, let us make the change of variables

\begin{eqnarray}
\label{VIA}
\Pi_f=\lambda_f(a_1a_2a_3);~~X^f=\hbox{ln}\Bigl({{a_f} \over {a_0}}\Bigr),
\end{eqnarray}

\noindent
where $a_0$ is a numerical constant of mass dimension $[a_0]=1$, and $(\hbox{det}A)=a_1a_2a_3\neq{0}$.  Equation (\ref{VIA}) imposes the following ranges on the configuration space $-\infty<\vert{X}^f\vert<\infty$, corresponding to 
${0}<\vert{a}_f\vert<\infty$.  The starting action (\ref{INSTANT711}) in terms of the new variables is given by 

\begin{eqnarray}
\label{THENEWACTION}
I_{Kin}=-{i \over G}\int{dt}\int_{\Sigma}d^3x\Bigl[\Pi_f\dot{X}^f-iNa_0^{3/2}e^{T/2}U\sqrt{\Pi_1\Pi_2\Pi_3}\Bigl({1 \over {\Pi_1}}+{1 \over {\Pi_2}}+{1 \over {\Pi_3}}\Bigr)\Bigr],
\end{eqnarray}

\noindent
where we have defined $T=X^1+X^2+X^3$.  The quantity $U$, which depends completely on spatial gradients of $X^f$, is as defined in Appendix A.  Equation (\ref{THENEWACTION}) is canonically well-defined and will form the basis of the reduced classical theory and its quantization.  Note that regarding $\Pi_f$ and $X^f$ in (\ref{VIA}) as the fundamental variables implies a symplectic two form

\begin{eqnarray}
\label{CANONOC}
\boldsymbol{\Omega}=-{i \over G}\int_{\Sigma}d^3x{\delta\Pi_f}\wedge{\delta{X}^f}=-{i \over G}\delta\Bigl(\int_{\Sigma}d^3x\Pi_f\delta{X}^f\Bigr)=\delta\boldsymbol{\theta},
\end{eqnarray}

\noindent
which is the exact variation of the canonical one form $\boldsymbol{\theta}$.  We will use (\ref{THENEWACTION}) as the starting point for formulation of the classical and quantum dynamics for GR on the 
kinematic phase space $\Omega_{Kin}$.

\subsection{Hamiltonian formalism}

Since (\ref{THENEWACTION}) already appears in first order form, we can directly read off from the canonical structure the following elementary Poisson brackets

\begin{eqnarray}
\label{CONJUGATE1}
\{X^f(x,t),\Pi_g(y,t)\}=iG\delta^f_g\delta^{(3)}(x,y),
\end{eqnarray}

\noindent
whence $\Pi_f$ is the momentum canonically conjugate to $X^f$.  The momentum conjugate to $N$ is given by

\begin{eqnarray}
\label{CONJUGATE2}
\Pi_N={{\delta{I}_{Kin}} \over {\delta\dot{N}}}=0,
\end{eqnarray}

\noindent
which implies the primary constraint $\Pi_N=0$.  Conservation of this constraint under time evolution leads to the secondary constraint

\begin{eqnarray}
\label{CONJUGATE3}
\dot{\Pi}_N=-{{\delta{I}_{Kin}} \over {\delta{N}}}=a_0^{3/2}e^{T/2}U\sqrt{\Pi_1\Pi_2\Pi_3}\Phi=0,
\end{eqnarray}

\noindent
where we have made the definition

\begin{eqnarray}
\label{BROOK}
\Phi={1 \over {\Pi_1}}+{1 \over {\Pi_2}}+{1 \over {\Pi_3}}.
\end{eqnarray}

\noindent
We must now check for the preservation of (\ref{CONJUGATE3}) under Hamiltonian evolution.  To carry this out we will need to evaluate Poisson brackets

\begin{eqnarray}
\label{BRACK}
\{H[M],H[N]\}
=\int_{\Sigma}d^3x\Bigl({{\delta{H}[M]} \over {\delta{X}^f}}{{\delta{H}[N]} \over {\delta{\Pi}_f}}
-{{\delta{H}[N]} \over {\delta{X}^f}}{{\delta{H}[M]} \over {\delta{\Pi}_f}}\Bigr)
\end{eqnarray}

\noindent
using the smeared Hamiltonian constraint, which is given by

\begin{eqnarray}
\label{CONJUGATE4}
H[N]=\int_{\Sigma}d^3xNa_0^{3/2}e^{T/2}U\sqrt{\Pi_1\Pi_2\Pi_3}\Phi.
\end{eqnarray}

\noindent
The functional derivative of (\ref{CONJUGATE4}) with respect to $\Pi_f$ is of the form

\begin{eqnarray}
\label{CONJUGATE5}
{{\delta{H}[N]} \over {\delta\Pi_f}}=N\Bigl(q_f\Phi+q\Bigl({1 \over {\Pi_f}}\Bigr)^2\Bigr),
\end{eqnarray}

\noindent
where $q$ and $q_f$ are functions on phase space, whose specific forms are not important for what follows.  The functional derivative with respect to $X^f$ is of the form

\begin{eqnarray}
\label{CONJUGATE6}
{{\delta{H}[N]} \over {\delta{X}^f}}=Q_fN\Phi+Q_{fi}\partial_i(QN\Phi)
\end{eqnarray}

\noindent
for some $Q$, $Q_f$ and $Q_{fi}$ which are phase space functions, again whose specific form is also not needed.  The spatial gradients in (\ref{CONJUGATE6}) originated from $U$ by integration of parts.\par
\indent
We will now compute the algebra of the Hamiltonian constraint $H$ 

\begin{eqnarray}
\label{CONJUGATE7}
\{H[M],H[N]\}=\int_{\Sigma}d^3xM\Bigl(q_f\Phi+q\Bigl({1 \over {\Pi_f}}\Bigr)^2\Bigr)\bigl(Q_fN\Phi+Q_{fi}\partial_i(QN\Phi)-N\leftrightarrow{M}.
\end{eqnarray}

\noindent
All terms which are proportional to $\Phi$ vanish on-shell on account of (\ref{BROOK}), which is implied by the Hamiltonian constraint.  So we need only consider terms from (\ref{CONJUGATE7}) of the form

\begin{eqnarray}
\label{CONJUGATE8}
\int_{\Sigma}d^3xMq\Bigl({1 \over {\Pi_f}}\Bigr)^2Q_{fi}\partial_i(QN\Phi)-N\leftrightarrow{M},
\end{eqnarray}

\noindent
and the only nontrivial contributions to (\ref{CONJUGATE8}) are due to the spatial gradients acting on the smearing functions $M$ and $N$.  This yields

\begin{eqnarray}
\label{CONJUGATE9}
\int_{\Sigma}d^3xqQ\Bigl({1 \over {\Pi_f}}\Bigr)^2Q_{fi}\bigl(M\partial_iN-N\partial_iM\bigr)\Phi.
\end{eqnarray}

\noindent
The result is that

\begin{eqnarray}
\label{CONJUGATE10}
\{H[M],H[N]\}=\{H[Q^i\bigl(M\partial_iN-N\partial_iM\bigr)\},
\end{eqnarray}

\noindent
where $Q^i=Q^i(X^f,\Pi_f)$ are phase space dependent structure functions.  The Poisson bracket of two Hamiltonian constraints $H$ on the phase space $\Omega_0=(X^f,\Pi_f)$ is proportional to a Hamiltonian constraint.  Therefore $H$ is first class and there are no second class constraints.  Since we started with a phase space of $2\times{3}=6$ degrees of freedom, the degrees of freedom per point subsequent to implementation of the Hamiltonian constraint are

\begin{eqnarray}
\label{DEGREESOF}
D.O.F.=2\times{3}-2\times{1}=4.
\end{eqnarray}

\noindent
With four phase space degrees of freedom per point, this shows that the reduced theory is not a topological field theory.

\section{Classical dynamics for $\Lambda=0$}

We will now formulate the classical dynamics on $\Omega_{Kin}$ for $\Lambda=0$.  For our starting action we will take the first order action given by

\begin{eqnarray}
\label{STARTING}
I_{Kin}={1 \over G}\int_{\Sigma}d^3x\Bigl(\Pi_f\dot{X}^f-iNa_0^{3/2}e^{T/2}U\sqrt{\Pi_1\Pi_2\Pi_3}\Phi\Bigr),
\end{eqnarray}

\noindent
where $U$, which contains spatial gradients of the configuration variables $X^f$, is as defined in Appendix A.  Also we have defined 

\begin{eqnarray}
\label{STARTING1}
\Phi={1 \over {\Pi_1}}+{1 \over {\Pi_2}}+{1 \over {\Pi_3}}.
\end{eqnarray}

\noindent
There are seven fields, $\Pi_f=(\Pi_1,\Pi_2,\Pi_3)$ which we require to be nonvanishing, $X^f=(X^1,X^2,X^3)$, and $N$ and we have defined $T=X^1+X^2+X^3$.  The Euler--Lagrange equations of motion from (\ref{STARTING}) are given by

\begin{eqnarray}
\label{STARTING2}
{d \over {dt}}\Bigl({{\delta{L}} \over {\delta\dot{N}}}\Bigr)={{\delta{I}_{Kin}} \over {\delta{N}}}.
\end{eqnarray}

\noindent
It is clear from the starting action (\ref{STARTING}) that the velocity $\dot{N}$ is absent.  Additionally, $N$ does not multiply a velocity, therefore it is an auxilliary field and (\ref{STARTING2}) yields

\begin{eqnarray}
\label{STARTING3}
a_0^{3/2}e^{T/2}U\sqrt{\Pi_1\Pi_2\Pi_3}\Phi=0.
\end{eqnarray}

\noindent
We require that $e^{T/2}U\sqrt{\Pi_1\Pi_2\Pi_3}$ be nonzero, hence (\ref{STARTING3}) reduces to

\begin{eqnarray}
\label{STARTING4}
\Phi={1 \over {\Pi_1}}+{1 \over {\Pi_2}}+{1 \over {\Pi_3}}=0,
\end{eqnarray}

\noindent
which is a constraint on the variables $\Pi_f$.  Note that this constraint is independent of the other variables $X^f$ and $N$.\par
\indent
The equation of motion for $X^f$ is given by

\begin{eqnarray}
\label{STARTING41}
{d \over {dt}}\Bigl({{\delta{I}_{Kin}} \over {\delta\dot{X}^f}}\Bigr)={{\delta{L}} \over {\delta{X}^f}},
\end{eqnarray}

\noindent
which is

\begin{eqnarray}
\label{STARTING42}
\dot{\Pi}_f=-Na_0^{3/2}e^{T/2}{{\delta{U}} \over {\delta{X}^f}}\{\sqrt{\Pi_1\Pi_2\Pi_3}\Phi\}.
\end{eqnarray}

\noindent
There are spatial gradients from $U$ which act on the terms in curly brackets.  But since these terms are proportional to $\Phi$, they vanish on solutions to (\ref{STARTING4}).  This implies that

\begin{eqnarray}
\label{STARTING43}
\Pi_f(x,t)=\Pi_f(x),
\end{eqnarray}

\noindent
which are arbitrary functions of position, independent of time.\par
\indent
To find the equations of motion for $\Pi_f$, we subtract a total time derivative ${d \over {dt}}(\Pi_fX^f)$ from the starting action (\ref{STARTING}) an obtain

\begin{eqnarray}
\label{STARTING5}
{d \over {dt}}\Bigl({{\delta{L}} \over {\delta\dot{\Pi}_f}}\Bigr)={{\delta{L}} \over {\delta\Pi_f}},
\end{eqnarray}

\noindent
which is 

\begin{eqnarray}
\label{STARTING6}
-\dot{X}^f=-Na_0^{3/2}e^{T/2}{{\delta(\Pi_1\Pi_2\Pi_3)^{1/2}} \over {\delta\Pi_f}}\Phi
-Na_0^{3/2}e^{T/2}U\sqrt{\Pi_1\Pi_2\Pi_3}\Bigl({{\delta\Phi} \over {\delta\Pi_f}}\Bigr).
\end{eqnarray}

\noindent
The first term on the right hand side of (\ref{STARTING6}) vanishes on account of (\ref{STARTING4}), and we are left with the following equations

\begin{eqnarray}
\label{STARTING7}
\dot{X}^1=-Na_0^{3/2}e^{T/2}U\sqrt{\Pi_1\Pi_2\Pi_3}\Bigl({1 \over {\Pi_1}}\Bigr)^2;\nonumber\\
\dot{X}^2=-Na_0^{3/2}e^{T/2}U\sqrt{\Pi_1\Pi_2\Pi_3}\Bigl({1 \over {\Pi_2}}\Bigr)^2;\nonumber\\
\dot{X}^3=-Na_0^{3/2}e^{T/2}U\sqrt{\Pi_1\Pi_2\Pi_3}\Bigl({1 \over {\Pi_3}}\Bigr)^2.
\end{eqnarray}

\noindent
It will be convenient to make the following definitions

\begin{eqnarray}
\label{STARTING8}
\eta=a_0^{3/2}\sqrt{\Pi_1\Pi_2\Pi_3}\Bigl(\Bigl({1 \over {\Pi_1}}\Bigr)^2+\Bigl({1 \over {\Pi_2}}\Bigr)^2+\Bigl({1 \over {\Pi_3}}\Bigr)^2\Bigr);\nonumber\\
\eta_f=a_0^{3/2}\sqrt{\Pi_1\Pi_2\Pi_3}\Bigl({1 \over {\Pi_f}}\Bigr)^2;~~\eta=\eta_1+\eta_2+\eta_3.
\end{eqnarray}

\noindent
where $\Pi_3=-{{\Pi_1\Pi_2} \over {\Pi_1+\Pi_2}}$ from (\ref{STARTING4}).  Then defining $T=X^1+X^2+X^3$, then (\ref{STARTING7}) is given by

\begin{eqnarray}
\label{STARTING8}
\dot{X}^f=\Bigl({{\eta_f} \over \eta}\Bigr)\dot{T};~~\dot{T}=-NUe^{T/2}\eta.
\end{eqnarray}

\noindent
We have to integrate the equation for $T$

\begin{eqnarray}
\label{STARTING9}
-e^{-T/2}\dot{T}=2{d \over {dt}}e^{-T/2}=NU\eta
\end{eqnarray}

\noindent
which yields

\begin{eqnarray}
\label{STARTING10}
e^{-T/2}=e^{-T_0/2}+{{\eta(x)} \over 2}\int^t_0N(x,t^{\prime})U(x,t^{\prime};T)dt^{\prime},
\end{eqnarray}

\noindent
where we have defined $T_0=T(x,0)$.  Equation (\ref{STARTING10}) is a nonlinear relation between $T$ and itself.  This can be written as

\begin{eqnarray}
\label{STARTING11}
T=\hbox{ln}\Bigl(e^{-T_0/2}+{{\eta(x)} \over 2}\int^t_0N(x,t^{\prime})U(x,t^{\prime};T)dt^{\prime}\Bigr)^{-2}.
\end{eqnarray}

\noindent
One may proceed from (\ref{STARTING11}) to perform a fixed point iteration procedure.  Define a sequence $T_n(x,t)$ where $T_0(x,t)=T_0$, and the following recursion relation holds

\begin{eqnarray}
\label{STARTING11}
T_{n+1}(x,t)=\hbox{ln}\Bigl(e^{-T_0/2}+{{\eta(x)} \over 2}\int^t_0N(x,t^{\prime})U(x,t^{\prime};T_n(x,t^{\prime}))dt^{\prime}\Bigr)^{-2}.
\end{eqnarray}

\noindent
For given initial data $X^f(x,0)$ on a 3 dimensional spatial hypersurface $\Sigma$ and a choice of the lapse function $N(x,t)$ through spacetime, if the iteration converges to a fixed point, then one has that 

\begin{eqnarray}
\label{STARTING12}
\hbox{lim}_{n\rightarrow\infty}T_n(x,t)=T(x,t).
\end{eqnarray}

\noindent
Integration of the first equation of (\ref{STARTING8}) yields the motion of $X^f$ 

\begin{eqnarray}
\label{STARTING13}
X^f(x,t)=X^f(x,0)+\Bigl({{\eta_f} \over \eta}\Bigr)T(x,t),
\end{eqnarray}

\noindent
with $T(x,t)$ given by (\ref{STARTING11}).  The variables $X^f$ evolve linearly with respect to $T$, seen as a time variable on configuration space $\Gamma$.\footnote{This seems to be the nearest gravitational analogy to the motion of a free particle in ordinary classical mechanics.}  The solutions for $X^f(x,t)$ in principle are directly 
constructible from (\ref{STARTING11}) and (\ref{STARTING12}), combined with the specification of boundary data $X^f(x,0)$.  Note that the solutions are labelled by two arbitrary functions of position $\Pi_1(x)$ and $\Pi_2(x)$.

\subsection{The spacetime metric}

\noindent
The spacetime metric is not a fundamental object and must be derived.  The fundamental objects are $X^f$, or alternatively the corresponding connection components which are given by exponentiation of (\ref{STARTING13}) 

\begin{eqnarray}
\label{FUNDAMENTAL}
a_f(x,t)=a_0\Bigl((\hbox{det}a(x,0)/a_0^3)^{-1/2}+{{\eta(x)} \over 2}\int^t_0N(x,t^{\prime})U(x,t^{\prime};T)dt^{\prime}\Bigr)^{-2\eta_f/\eta}.
\end{eqnarray}

\noindent
Equation (\ref{FUNDAMENTAL}) provides the explicit time variation for the diagonal connection in the reduced full theory.  Taking the product over $i=1,2,3$ one finds that for $t=0$ the condition $\hbox{det}a=\hbox{det}a(x,0)$ is satisfied, which can be chosen arbitrarily on the initial spatial hypersurface $\Sigma_0$.  One must then choose the lapse function $N(x,t)$ to specify the manner in which the boundary data becomes evolved for $t>0$.  The solutions are labelled by the conjugate momenta $\Pi_f$ as encoded in $\eta_f/\eta$.  Equation (\ref{FUNDAMENTAL}) can also be written as

\begin{eqnarray}
\label{FUNDAMENTAL1}
a_f(x,t)=\Bigl({{\hbox{det}a(x,t)} \over {\hbox{det}a(x,0)}}\Bigr)^{\eta_f/\eta}=a_0e^{(\eta_f/\eta)T},
\end{eqnarray}

\noindent
whence the variables evolve with respect to $\hbox{det}a$, seen as a time variable on configuration space.  We will illustrate the construction of the metric for a simple example where the spatial gradients are zero.  Recall in the original Ashtekar variables that the contravariant 3-metric $h^{ij}$ is given by

\begin{eqnarray}
\label{THEMET}
hh^{ij}=\widetilde{\sigma}^i_a\widetilde{\sigma}^j_a\longrightarrow
{h}^{ij}=(\hbox{det}\widetilde{\sigma})^{-1}\widetilde{\sigma}^i_a\widetilde{\sigma}^j_a.
\end{eqnarray}

\noindent
The covariant form on the phase space $\Omega_{Inst}$ is given by

\begin{eqnarray}
\label{THEMET1}
h_{ij}=(\hbox{det}\Psi)\Psi^{-1}_{ae}\Psi^{-1}_{af}(B^{-1})^e_i(B^{-1})^f_j(\hbox{det}B).
\end{eqnarray}

\noindent
Restricted to the subspace of diagonal connection variables, which admit the proper canonical relation to the densitized eigenvalues of the CDJ matrix $\lambda_f$, this is given by

\begin{displaymath}
h_{ij}=(\lambda_1\lambda_2\lambda_3)
\left(\begin{array}{ccc}
(a_1/\lambda_1)^2 & 0 & 0\\
0 & (a_2/\lambda_2)^2 & 0\\
0 & 0& (a_3/\lambda_3)^2\\
\end{array}\right)
\end{displaymath}

\noindent
which upon the subsitution $\lambda_i=\Pi_i(\hbox{det}a)^{-1}$ yields

\begin{eqnarray}
\label{METRICC}
h_{ij}=\delta_{ij}(\Pi_1\Pi_2\Pi_3)\Bigl({{a_0^3} \over {\hbox{det}a}}\Bigr)\Bigl({{a_j} \over {\Pi_j}}\Bigr)^2
\end{eqnarray}

\noindent
with $a_j$ given by (\ref{FUNDAMENTAL}).  For simplicity consider the case where the variables are independent of spatial position and depend only on time.  Then $\Pi_i$ are numerical constants, $a_i(x,t)=a_i(t)$, and 
moreover $U=1$.  As as special case, take $a_i(x,0)=a_0$, and take $N(x,t)=2$, namely a constant lapse.  Then the metric evolves in time via

\begin{eqnarray}
\label{METRICC1}
ds^2=dt^2+\delta_{ij}\Bigl({{\Pi_1\Pi_2\Pi_3} \over {\Pi_j^2}}\Bigr)(1+\eta{t})^{2(1-\eta_j/\eta)}dx^idx^j,
\end{eqnarray}

\noindent
which has the same form as the Kasner solution, with a re-definition of variables.  One may compute the initial volume of the universe

\begin{eqnarray}
\label{METRICC2}
Vol(\Sigma_0)=\int_{\Sigma}d^3x\sqrt{h}=l^3\Bigl({{\Pi_1\Pi_2\Pi_3} \over {\hbox{det}a(0)}}\Bigr)=l^3(\hbox{det}a(0))^{-1}\Bigl({{(\Pi_1\Pi_2)^2} \over {\Pi_1+\Pi_2}}\Bigr)
\end{eqnarray}

\noindent
at $t=0$, where $l$ is a characteristic length scale of the universe from integration over minisuperspace.  Note that this volume is labelled by two arbitrary constants $\Pi_1$ and $\Pi_2$ which determine the algebraic classification of the spacetime, as well as $\hbox{det}a(0)$.  This provides a physical interpretation for $\hbox{det}a$ in terms of metric variables.  A more in-depth analysis of minisuperspace, as well as a generalization of the above procedure to the full theory, is reserved for a separate paper.

\newpage

\section{Quantization and Hilbert space structure for vanishing cosmological constant}

We now proceed to the quantum theory on the kinematic phase space.  We have already eliminated the Gauss' law and diffeomorphism constraints, leaving behind a Dirac consistent phase space which admits a canonical formulation and classical dynamics.  This implies that we may proceed to the quantum theory by promoting the dynamical variables to quantum operators $X^f\rightarrow\hat{X}^f$ and $\Pi_f\rightarrow\hat{\Pi}_f$, and Possion 
brackets (\ref{CONJUGATE1}) to commutators

\begin{eqnarray}
\label{CONJUGATE11}
\bigl[\hat{X}^f(x,t),\hat{\Pi}_g(y,t)\bigr]=(\hbar{G})\delta^f_g\delta^{(3)}(x,y).
\end{eqnarray}

\noindent
The operators in the functional Schr\"odinger representation act respectively by multiplication and by functional differentiation of a wavefunctional

\begin{eqnarray}
\label{CONJUGATE12}
\hat{X}^f(x,t)\boldsymbol{\psi}=X^f(x,t)\boldsymbol{\psi};\nonumber\\
\hat{\Pi}_f(x,t)\boldsymbol{\psi}=(\hbar{G}){\delta \over {\delta{X}^f(x,t)}}\boldsymbol{\psi}.
\end{eqnarray}

\noindent
Note that the following wavefunctionals are eigenstates of $\hat{\Pi}_f$

\begin{eqnarray}
\label{CONJUGATE13}
\boldsymbol{\psi}_{\lambda}[X]=\hbox{exp}\Bigl[(\hbar{G})^{-1}\int_{\Sigma}d^3x\widetilde{\lambda}_f(x)X^f(x,t)\Bigr],
\end{eqnarray}

\noindent
where $\widetilde{\lambda}_f(x)$ are arbitrary continuous functions of position, which do not contain any functional dependence on $X^f(x,t)$.  We will see that these play the role of labels for the state.  The following action ensues for the momentum operator

\begin{eqnarray}
\label{CONJUGATE14}
\hat{\Pi}_f(x,t)\boldsymbol{\psi}_{\lambda}[X]=\widetilde{\lambda}(x)\boldsymbol{\psi}_{\lambda}[X].
\end{eqnarray}

\noindent
We will now search for states $\boldsymbol{\psi}\in{Ker}\{\hat{H}\}$.  But prior to quantization let us put the smeared constraint into polynomial form

\begin{eqnarray}
\label{CONJUGATE15}
H[N]=\int_{\Sigma}d^3xNa_0^{3/2}e^{T/2}U(\Pi_1\Pi_2\Pi_3)^{-1/2}\bigl(\Pi_1\Pi_2+\Pi_2\Pi_3+\Pi_3\Pi_1\bigr).
\end{eqnarray}

\noindent
To obtain a nontrivial solution it suffices for the operator in brackets in (\ref{CONJUGATE15}) upon quantization to annihilate the state for each $x$.  Hence

\begin{eqnarray}
\label{CONJUGATE16}
\bigl(\hat{\Pi}_1(x)\hat{\Pi}_2(x)+\hat{\Pi}_2(x)\hat{\Pi}_3(x)+\hat{\Pi}_3(x)\hat{\Pi}_1(x)\bigr)\boldsymbol{\psi}_{\lambda}[X]=0~~\forall{x}\nonumber\\
\longrightarrow\Bigl(\widetilde{\lambda}_1(x)\widetilde{\lambda}_2(x)+\widetilde{\lambda}_2(x)\widetilde{\lambda}_3(x)+\widetilde{\lambda}_3(x)\widetilde{\lambda}_1(x)\Bigr)
\boldsymbol{\psi}_{\lambda}[X]=0~~\forall{x}.
\end{eqnarray}

\noindent
This leads to the dispersion relation

\begin{eqnarray}
\label{CONJUGATE17}
\widetilde{\lambda}_3=-\Bigl({{\widetilde{\lambda}_1\widetilde{\lambda}_2} \over {\widetilde{\lambda}_1+\widetilde{\lambda}_2}}\Bigr)~~\forall{x}.
\end{eqnarray}

\noindent
Conventionally in quantum field theory, when there are products of momenta evaluated at the same point a regularization procedure is needed to obtain a well-defined action on states.  However, there exist states for which the 
action of (\ref{CONJUGATE6}), is already well-defined without the need for regularization, namely plane wave-type states annihilated by $\hat{\Phi}$.  These are states for which the momenta are functionally independent of the configuration variables and act as labels.  The solution is given by\footnote{We use the tilde notation to distinguish $\widetilde{\lambda}_f$, the eigenvalue of $\hat{\Pi}_f$ on $\boldsymbol{\psi}$, from the 
(undensitized) eigenvalues $\lambda_f$ of $\Psi_{(ae)}$.  Since $\Pi_f=\lambda_f(\hbox{det}a)$ at the classical level, then $\widetilde{\lambda}_f$ can be seen as a `densitized' version of $\lambda_f$.  We do not include the tilde in the specification of the state $\bigl\vert\lambda_1,\lambda_2\bigr>$, since it would be redundant owing to the invariance of $\Phi$ under rescaling of $\lambda_f$ for $\Lambda=0$.}

\begin{eqnarray}
\label{QUANTIZATION4}
\boldsymbol{\psi}_{\lambda_1,\lambda_2}[X(x)]=\hbox{exp}\Bigl[(\hbar{G})^{-1}\sum_f\widetilde{\lambda}_f(x)X^f(x)\Bigr)\Bigr]\Biggl\vert_{\lambda_3=-\lambda_1\lambda_2/(\lambda_1+\lambda_2)}
\end{eqnarray}

\noindent
for each $x\in\Sigma$.  Hence $\bigl\vert\lambda\bigr>=\bigl\vert\lambda_1,\lambda_2\bigr>\in{Ker}\{\hat{\Phi}\}$ defines a Hilbert space of states annihilated by the Hamiltonian constraint, labelled by $\lambda_1$ and $\lambda_2$, once the measure of normalization has been defined.  The full Hilbert space consists of a direct product of the Hilbert spaces $\forall{x}\in\Sigma$, since (\ref{CONJUGATE17}) must be satisfied independently at each point $x$.  If one regards each spatial hypersurface $\Sigma$ as a lattice of finite lattice spacing $\textbf{x}_{n+1}-\textbf{x}_n=\Delta{x}$, then

\begin{eqnarray}
\label{QUANTIZATION5}
\boldsymbol{H}=\bigotimes_{\textbf{x}_n}\boldsymbol{H}(\textbf{x}_n)\longrightarrow\boldsymbol{\psi}_{\lambda_1\lambda_2}\sim\prod_{\textbf{x}_n}\boldsymbol{\psi}_{\lambda_1\lambda_2}(\textbf{x}_n).
\end{eqnarray}

\noindent
In the continuum limit $\Delta{x}\rightarrow{0}$, the product in (\ref{QUANTIZATION5}) goes to a Riemannian integral 

\begin{eqnarray}
\label{QUANTIZATION6}
\boldsymbol{\psi}_{\lambda_1,\lambda_2}[X]
=\hbox{exp}\biggl[(\hbar{G})^{-1}\int_{\Sigma}d^3x\Bigl(\widetilde{\lambda}_1X^1+\widetilde{\lambda}_2X^2
-\Bigl({{\widetilde{\lambda}_1\widetilde{\lambda}_2} \over {\widetilde{\lambda}_1+\widetilde{\lambda}_2}}\Bigr)X^3\Bigr)\biggr].
\end{eqnarray}

\noindent
Equation (\ref{QUANTIZATION6}) solves the quantum Hamiltonian constraint by construction.  The momentum labels $(\lambda_1,\lambda_2)$ correspond to two functions of spatial position $\textbf{x}\in\Sigma$.\par  
\indent

\subsection{Measure on the Hilbert space }

\noindent
To formalize the Hilbert space structure we need square integrable wavefunctions for solutions to the constraints, which requires the specification of a measure for normalization.  If all variables were real, as for spacetimes of Euclidean signature, one would be able to use delta-functional normalizable wavefunctions.    

\begin{eqnarray}
\label{HEEL1}
D\mu_{Eucl}(X)=\prod_{\textbf{x}}\delta{X}^1(x)\delta{X}^2(x)\delta{X}^3(x).  
\end{eqnarray}

\noindent
In (\ref{HEEL1}) $X^f$ is real and on the replacement $\lambda_f\rightarrow{i}\lambda_f$, we have

\begin{eqnarray}
\label{HEEL2}
\bigl<\psi_{\lambda}\bigl\vert\psi_{\zeta}\bigr>_{Eucl}
=D\mu_{Eucl}(\xi)\hbox{exp}\Bigl[-i(\hbar{G})^{-1}\int_{\Sigma}d^3x\widetilde{\lambda}_f(x)X^f(x)\Bigr]\nonumber\\
\hbox{exp}\Bigl[i(\hbar{G})^{-1}\int_{\Sigma}d^3x\widetilde{\zeta}_f(x)X^f(x)\Bigr]
=\prod_{\textbf{x}}\prod_f\delta\bigl(\widetilde{\lambda}_f(\textbf{x})-\widetilde{\zeta}_f(\textbf{x})\bigr),
\end{eqnarray}

\noindent
or that two states are orthogonal unless their CDJ matrix eigenvalues are identical at each point $\textbf{x}\in\Sigma$.  This can be written more compactly as

\begin{eqnarray}
\label{HEEL3}
\bigl<\psi_{\lambda}\bigl\vert\psi_{\zeta}\bigr>_{Eucl}=\int_{\Gamma}{D}\mu_{Eucl}(\xi)e^{-i(\hbar{G})^{-1}\widetilde{\lambda}\cdot{X}}e^{i(\hbar{G})^{-1}\widetilde{\zeta}\cdot{X}}=\delta_{\lambda\zeta}.
\end{eqnarray}
 
\indent
For spacetimes of Lorentzian signature, the variables are in general complex and a Euclidean measure does not produce normalizable wavefunctions.  One may then rather use a Gaussian measure to ensure square integrability for the basis wavefunctions in this case.  This Gaussian measure is given by

\begin{eqnarray}
\label{HEEL4}
D\mu_{Lor}(\overline{X},X)=\bigotimes_{\textbf{x}}\nu^{-1}\delta{\xi}e^{-\nu^{-1}\overline{X}\cdot{X}}\nonumber\\
=\prod_{\textbf{x},f}\delta{X}^f\hbox{exp}\Bigl[-\nu^{-1}\int_{\Sigma}d^3x\overline{X}_f(x)X^f(x)\Bigr],
\end{eqnarray}

\noindent
where $\nu$ is a numerical constant with mass dimensions $[\nu]=-3$, needed to make the argument of the exponential dimensionless.  The inner product of two un-normalized states is now given by

\begin{eqnarray}
\label{HEEL5}
\bigl<\lambda\bigl\vert\zeta\bigr>_{Lor}
=\prod_{\textbf{x},i}\int_{\Gamma}\nu^{\zeta(0)}\delta{X}^f\hbox{exp}\Bigl[-\nu^{-1}\int_{\Sigma}d^3x\overline{X}_f(x){X}^f(x)\Bigr]\nonumber\\
\hbox{exp}\Bigl[(\hbar{G})^{-1}\int_{\Sigma}d^3x\widetilde{\lambda}^{*}_f(x)\overline{\xi}_f(x)\Bigr]
\hbox{exp}\Bigl[(\hbar{G})^{-1}\int_{\Sigma}d^3x\widetilde{\zeta}_f(x){X}^f(x)\Bigr]\nonumber\\
=\hbox{exp}\Bigl[\nu(\hbar{G})^{-2}\int_{\Sigma}d^3x\widetilde{\lambda}^{*}_f(x)\widetilde{\zeta}_f(x)\Bigr].
\end{eqnarray}

\noindent
A necessary condition for the wavefunction to be normalizable, as for the inner product to exist, is that the functions $\widetilde{\lambda}_i(x)$ and $\widetilde{\zeta}_i(x)$ be square 
integrable.  In shorthand notation, (\ref{HEEL5}) can be written as

\begin{eqnarray}
\label{HEEL6}
\bigl<\lambda\bigl\vert\zeta\bigr>_{Lor}=\int_{\Gamma}{D}\mu_{Lor}(\overline{X},X)e^{(\hbar{G})^{-1}\widetilde{\lambda}^{*}\cdot\overline{X}}e^{(\hbar{G})^{-1}\widetilde{\zeta}\cdot{X}}=
e^{\nu(\hbar{G})^{-2}\widetilde{\lambda}^{*}\cdot\widetilde{\zeta}}.
\end{eqnarray}

\noindent
Note how the balance of the mass dimensions is ensured in spite of the existence of infinite dimensional spaces.\footnote{The dimensionful constant $\nu$ remains a parameter of the theory.  One may think that such a measure cannot exist on infinite dimensional spaces unless $\nu=1$ with $[\nu]=0$.  But we have rescaled the measure by the same factor of $\nu^{\zeta(0)}$ to cancel out these factors arising from the Gaussian integral.}  The norm of a state is given by

\begin{eqnarray}
\label{DELTAA2}
\bigl<\lambda\bigr\vert\lambda\bigr>
=\int{D\mu}_{Lor}(\xi,\overline{\xi})e^{(\hbar{G})^{-1}\widetilde{\lambda}^{*}\cdot\overline{\xi}}e^{(\hbar{G})^{-1}\widetilde{\lambda}\cdot\xi}
=e^{\nu(\hbar{G})^{-2}\widetilde{\lambda}^{*}\cdot\widetilde{\lambda}},
\end{eqnarray}

\noindent
and we define the normalized wavefunction by

\begin{eqnarray}
\label{DELTAA3}
\bigl\vert\boldsymbol{\psi}_{\lambda}\bigr>=e^{-\nu(\hbar{G})^{-2}\widetilde{\lambda}^{*}\cdot\widetilde{\lambda}}\bigl\vert\lambda\bigr>.
\end{eqnarray}

\noindent
The overlap of two states in the Lorentzian measure is given by

\begin{eqnarray}
\label{HEEL7}
\bigl\vert\bigl<\psi_{\lambda}\bigl\vert\psi_{\zeta}\bigr>_{Lor}\bigr\vert=\hbox{exp}\Bigl[-\nu(\hbar{G})^{-2}\int_{\Sigma}d^3x\bigl\vert\widetilde{\lambda}_i(x)-\widetilde{\zeta}_i(x)\vert^2\Bigr].
\end{eqnarray}

\noindent
where

\begin{eqnarray}
\label{DELTAA4}
\widetilde{\lambda}_3=-\Bigl({{\widetilde{\lambda}_1\widetilde{\lambda}_2} \over {\widetilde{\lambda}_1+\widetilde{\lambda}_2}}\Bigr);
~~\widetilde{\zeta}_3=-\Bigl({{\widetilde{\zeta}_1\widetilde{\zeta}_2} \over {\widetilde{\zeta}_1+\widetilde{\zeta}_2}}\Bigr).
\end{eqnarray}

\noindent
There is always a nontrivial overlap between any two states corresponding to different functions for the eigenvalues.\footnote{It is shown in \cite{ITAI} that the eigenvalues of $\Psi_{ae}$ encode the Petrov classification of spacetime, since $\Psi_{ae}$ is the antiself-dual part of the Weyl curvature tensor.  This classification is independent of coordinates and of tetrad frames.}

\subsection{Expectation values and observables}

\noindent
The expectation value of the configuration variable $X^f$ is given by

\begin{eqnarray}
\label{HEEL8}
\bigl<\psi_{\lambda}\bigl\vert\hat{X}^f(x)\bigr\vert\psi_{\zeta}\bigr>_{Lor}
=\prod_{\textbf{x},i}\int_{\Gamma}\nu^{\zeta(0)}\delta{X}^f\hbox{exp}\Bigl[-\nu^{-1}\int_{\Sigma}d^3x\overline{X}_f(x)X^f(x)\Bigr]\nonumber\\
\hbox{exp}\Bigl[(\hbar{G})^{-1}\int_{\Sigma}d^3x\widetilde{\lambda}^{*}_i(x)\overline{X}_f(x)\Bigr]
\Bigl(X^f(x)\hbox{exp}\Bigl[(\hbar{G})^{-1}\int_{\Sigma}d^3x\widetilde{\zeta}_f(x)X^f(x)\Bigr]\Bigr).\nonumber\\
\end{eqnarray}

\noindent
By replacing multiplication by $X^f$ with functional differentiation with respect to $\widetilde{\zeta}_f$, one may simplify the matrix element to

\begin{eqnarray}
\label{HEEL9}
\bigl<\psi_{\lambda}\bigl\vert\hat{X}^f(x)\bigr\vert\psi_{\zeta}\bigr>_{Lor}
=\prod_{\textbf{x},i}\int_{\Gamma}\nu^{\zeta(0)}\delta\xi_i\hbox{exp}\Bigl[-\nu^{-1}\int_{\Sigma}d^3x\overline{X}_f(x)X^f(x)\Bigr]\nonumber\\
\hbox{exp}\Bigl[(\hbar{G})^{-1}\int_{\Sigma}d^3x\widetilde{\lambda}^{*}_i(x)\overline{X}_f(x)\Bigr]
\Bigl({\delta \over {\delta\overline{\zeta}_i(x)}}\hbox{exp}\Bigl[(\hbar{G})^{-1}\int_{\Sigma}d^3x\widetilde{\zeta}_f(x)X^f(x)\Bigr]\Bigr),
\end{eqnarray}

\noindent
whereupon commuting the functional derivative outside the integral we obtain

\begin{eqnarray}
\label{HEEL10}
\bigl<\psi_{\lambda}\bigl\vert\hat{X}^f(x)\bigr\vert\psi_{\zeta}\bigr>_{Lor}
={\delta \over {\delta\widetilde{\zeta}_f(x)}}\Bigl(\hbox{exp}\Bigl[\nu(\hbar{G})^{-2}\int_{\Sigma}d^3x\widetilde{\lambda}^{*}_f(x)\widetilde{\zeta}_f(x)\Bigr]\Bigr)\nonumber\\
=\nu(\hbar{G})^{-2}\widetilde{\lambda}^{*}_f(x)\hbox{exp}\Bigl[\nu(\hbar{G})^{-2}\int_{\Sigma}d^3x\widetilde{\lambda}^{*}_f(x)\widetilde{\zeta}_f(x)\Bigr]\nonumber\\
=\bigl(\nu(\hbar{G})^{-2}\widetilde{\lambda}^{*}_f(x)\bigr)\bigl<\psi_{\lambda}\bigl\vert\psi_{\zeta}\bigr>_{Lor}.
\end{eqnarray}

\noindent
Going through a similar analysis for various operators, one obtains

\begin{eqnarray}
\label{HEEL11}
\bigl<\psi_{\lambda}\bigl\vert\hat{\overline{X}}_f(x)\bigr\vert\psi_{\zeta}\bigr>_{Lor}
=\bigl(\nu(\hbar{G})^{-2}\widetilde{\zeta}_f(x)\bigr)\bigl<\psi_{\lambda}\bigr\vert\psi_{\zeta}\bigr>_{Lor}
\end{eqnarray}

\begin{eqnarray}
\label{HEEL12}
\bigl<\psi_{\lambda}\bigl\vert\hat{\Pi}_f(x)\bigr\vert\psi_{\zeta}\bigr>_{Lor}
=\bigl<\psi_{\lambda}\bigl\vert(\hbar{G}){\delta \over {\delta{X}^f(x)}}\bigr\vert\psi_{\zeta}\bigr>_{Lor}
=\widetilde{\zeta}_f(x)\bigl<\psi_{\lambda}\bigr\vert\psi_{\zeta}\bigr>_{Lor}
\end{eqnarray}

\noindent
as well as

\begin{eqnarray}
\label{HEEL13}
\bigl<\psi_{\lambda}\bigl\vert{\delta \over {\delta\overline{X}_f}}\bigr\vert\psi_{\zeta}\bigr>_{Lor}
=(\hbar{G})^{-1}\widetilde{\lambda}^{*}_f(x)\bigl<\psi_{\lambda}\bigl\vert\psi_{\zeta}\bigr>_{Lor}.
\end{eqnarray}

\noindent
Hence, with respect to the Lorentzian measure one has, schematically,

\begin{eqnarray}
\label{HEEL14}
{\delta \over {\delta{X}^f}}\sim\hbar{G}\nu^{-1}\overline{X}_f;~~
{\delta \over {\delta\overline{X}_f}}\sim\hbar{G}\nu^{-1}{X}^f.
\end{eqnarray}

\noindent
This property of the infinite generalization of a Bargmann-like representation, combined with generating functional techniques, enables an explicit calculation of the matrix element of any observable $O$

\begin{eqnarray}
\label{HEEL15}
\bigl<\psi_{\lambda}\bigl\vert\hat{O}[\widetilde{X}_f;\widetilde{\lambda}_f]\bigr\vert\psi_{\zeta}\bigr>_{Lor}
=O[\nu(\hbar{G})^2\widetilde{\lambda}^{*}_f;\nu\widetilde{\lambda}_f]\bigl<\psi_{\lambda}\vert\psi_{\zeta}\bigr>_{Lor}.
\end{eqnarray}

\noindent
Hence, the existence of a function $O$ signifies the existence the expectation value or matrix element corresponding to $O$.

\section{Canonical equivalence to the Ashtekar variables}

\noindent
We have provided a direct map from the nondegenerate sector of the full phase space of the Ashtekar variables $\Omega_{Ash}=(\widetilde{\sigma}^i_a,A^a_i)$ to the phase space $\Omega_{Inst}=(\Psi_{ae},A^a_i)$ using the 
CDJ Ansatz $\widetilde{\sigma}^i_a=\Psi_{ae}B^i_e$.  By implementation of the Gauss' law and diffeomorphism constraints we have reduced $\Omega_{Inst}$ to $\Omega_{Kin}$, its kinematic phase space where we have computed the Hamiltonian dynamics.  Subsequently, we have performed a quantization of $\Omega_{Kin}$, obtaining a Hilbert space of states solving the quantum Hamiltonian constraint for $\Lambda=0$.  In this section we will demonstrate canonical equivalence to the Ashtekar theory which should imply that our results extend to certain regimes of the Ashtekar theory.  Note that both theories at the unconstrained level share in common the Ashtekar connection $A^a_i$ as the configuration space variable.  In what follows we will exploit the preservation of this property at all levels of reduction sequence.\par
\indent
We will prove, using the unconstrained Ashtekar theory as a starting point, that the map to $\Omega_{Kin}$ requires as a necessary and sufficient condition the implementation of the kinematic initial value constraints.  The canonical commutation relations for the Ashtekar variables are given by

\begin{eqnarray}
\label{THECOMMUTATION}
\bigl[A^a_i(x),\widetilde{\sigma}^j_b(y)\bigr]=\delta^a_b\delta^j_i\delta^{(3)}(x,y),
\end{eqnarray}

\noindent
where we have omitted the time dependence in order to avoid cluttering up the notation.  Let us now substitute the CDJ Ansatz $\widetilde{\sigma}^i_a=\Psi_{ae}B^i_e$ into (\ref{THECOMMUTATION})

\begin{eqnarray}
\label{THECOMMUTATION1}
\bigl[A^a_i(x),\Psi_{be}(y)B^j_e(y)\bigr]=\delta^a_b\delta^j_i\delta^{(3)}(x,y).
\end{eqnarray}

\noindent
We will now multiply (\ref{THECOMMUTATION1}) by $A^c_j(y)$ in the following form

\begin{eqnarray}
\label{THECOMMUTATION2}
\bigl[A^a_i(x),\Psi_{be}(y)B^j_e(y)A^c_j(y)\bigr]=\delta^a_bA^c_i(y)\delta^{(3)}(x,y),
\end{eqnarray}

\noindent
which is allowed since $[A^a_i,A^c_j]=0$ for the Ashtekar connection.  Define the magnetic helicity density matrix $C_{ce}=A^b_jB^j_e$, written in component form as

\begin{eqnarray}
\label{THECOMMUTATION3}
C_{ce}=\epsilon^{ijk}A^c_i\partial_jA^e_k+\delta_{ce}(\hbox{det}A),
\end{eqnarray}

\noindent
which has a diagonal part free of spatial gradients and an off-diagonal part containing spatial gradients.  Then the commutation relations read

\begin{eqnarray}
\label{THECOMMUTATION4}
\bigl[A^a_i(x),\Psi_{be}(y)C_{ce}(y)\bigr]=\delta^a_bA^c_i(y)\delta^{(3)}(x,y).
\end{eqnarray}

\noindent
The kinematic configuration space $\Gamma_{Kin}$ must have three degrees of freedom per point.\footnote{This is nine total degrees of freedom, minus three corresponding to $G_a$, and minus three corresponding to $H_i$.}  Let us choose for these D.O.F. to be the three diagonal elements $A^a_i=\delta^a_iA^a_a$.  Then we can set $a=i$ in (\ref{THECOMMUTATION4}) to obtain

\begin{eqnarray}
\label{THECOMMUTATION5}
\bigl[A^a_a(x),\Psi_{be}(y)C_{ce}(y)\bigr]=\delta^a_bA^c_a(y)\delta^{(3)}(x,y).
\end{eqnarray}

\noindent
Since $A^a_i$ is diagonal by supposition, then the only nontrivial contribution to (\ref{THECOMMUTATION5}) occurs for $a=c$.  Since $a=b$ also is the only nontrivial contribution, it follows that $b=c$ as well.  Hence the commutation relations for diagonal connection are given by

\begin{eqnarray}
\label{THECOMMUTATION6}
\bigl[A^a_a(x),\Psi_{be}(y)C_{be}(y)\bigr]=\delta^a_b\delta{A}^b_b(y)\delta^{(3)}(x,y).
\end{eqnarray}

\noindent
Substituting (\ref{THECOMMUTATION3}) subject to a diagonal connection into (\ref{THECOMMUTATION6}) we have

\begin{eqnarray}
\label{THECOMMUTATION7}
\sum_{e=1}^3\bigl[A^a_a(x),\Psi_{be}(y)\delta_{be}(\hbox{det}A)\bigr]\nonumber\\
+\sum_{e=1}^3\bigl[A^a_a(x),\Psi_{be}(y)\epsilon^{bje}A^b_b\partial_jA^e_e\bigr]=\delta^a_bA^b_b(y)\delta^{(3)}(x,y),
\end{eqnarray}

\noindent
which has split up into two terms.  We have been explicit in putting in the summation symbol to indicate that $e$ is a dummy index, while $a$ and $b$ are not.  There are two cases to consider, $e=b$ and $e\neq{b}$.  For $e\neq{b}$ the first term of (\ref{THECOMMUTATION7}) vanishes, leaving remaining the second term.  Since the right hand side stays the same, then this would correspond to the commutation relations for a CDJ matrix whose diagonal components are zero.  For the second possibility $e=b$ the second term of (\ref{THECOMMUTATION7}) vanishes while the first term survives, with the right hand side the same as before.  This case occurs only if the CDJ matrix $\Psi_{ae}$ is diagonal.  Let us choose $\Psi_{ae}=Diag(\lambda_1,\lambda_2,\lambda_3)$ as the diagonal matrix of eigenvalues,\footnote{This places one into the intrinsic $SO(3,C)$ frame.  Note that we may regard the Gauss' law constraint $G_a$ as already implemented in this frame, since it is a map from $\lambda_f$ to the $SO(3,C)$ angles $\vec{\theta}$ and not a constraint on $\lambda_f$.} then (\ref{THECOMMUTATION7}) reduces to

\begin{eqnarray}
\label{THECOMMUTATION8}
\bigl[A^a_a(x),\lambda_b(y)(\hbox{det}A(y))\bigr]=\delta^a_bA^a_a(y)\delta^{(3)}(x,y).
\end{eqnarray}

\noindent
The conclusion is that in order for (\ref{THECOMMUTATION8}) to have arisen from (\ref{THECOMMUTATION}), that: (i) The antisymmetric part of $\Psi_{ae}$ must be zero, namely, the diffeomorphism constraint must be satisfied. (ii) The symmetric off-diagonal part of $\Psi_{ae}$ is not part of the commutation relations on the diffeomorphism invariant phase space $\Omega_{diff}$.  Given the eigenvalues $\lambda_f$ on this space, the Gauss' law constraint can be solved separately from the quantization process.  The choice of a diagonal connection $A^a_a$ on $\Omega_{Kin}$ is consistent with the implementation of the kinematic constraints, which means that only the Hamiltonian constraint is necessary to obtain the physical phase space $\Omega_{Phys}$.\par
\indent
Equation (\ref{THECOMMUTATION8}) are not canonical commutation relations owing to the field-dependence on the right hand side.\footnote{While (\ref{THECOMMUTATION8}) are not canonical commutation relations, they are affine commutation relations which serve as an intermediate step in the formulation of canonical commutation relations.  Affine commutation relations have been used by Klauder in \cite{KLAUDER} in the affine quantum gravity programme, and are viable as well in the present case.}  However, they can be transformed into canonical commutation relations using the following change of variables $A^a_a=a_0e^{X^a}$ for $a=1,2,3$.  This yields

\begin{eqnarray}
\label{THECOMMUTATION911}
\bigl[e^{X^a(x)},\lambda_b(y)(\hbox{det}A(y))\bigr]=e^{X^a(x)}\bigl[X^a(x),\lambda_b(y)(\hbox{det}A(y))\bigr]=\delta^a_be^{X^a(y)}\delta^{(3)}(x,y).
\end{eqnarray}

\noindent
Since the only nontrivial contribution to (\ref{THECOMMUTATION911}) comes from $x=y$, we can cancel the pre-factor of $e^{X^a}$ from both sides.  Defining densitized eigenvalues $\Pi_b=\lambda_b(\hbox{det}A)$ as the fundamental momentum space variables, we have that the canonical version of (\ref{THECOMMUTATION8}) is given by

\begin{eqnarray}
\label{THECOMMUTATION10}
\bigl[X^a(x),\Pi_b(y)\bigr]=\delta^a_b\delta^{(3)}(x,y),
\end{eqnarray}

\noindent
The coordinate ranges are $\infty<\vert{X}^f\vert<\infty$, which corresponds to $0<\vert{A}^f_f\vert<\infty$, which is a subset of the latter.  To utilize the full range of $A^a_i$, which includes the degenerate 
cases, one may instead use (\ref{THECOMMUTATION8}).  We have shown that $\Omega_{Kin}$ of the instanton representation admits a cotangent bundle structure with diagonal connection $A^a_a(x)$.  It happens from (\ref{THECOMMUTATION}) that $A^a_a(x)$ is canonically conjugate to $\widetilde{\sigma}^a_a(x)$.  Since the instanton representation maps to the Ashtekar formalism and vice versa on the unreduced phase space for nondegenerate $B^i_a$, it 
follows that (\ref{THECOMMUTATION10}) corresponds as well to the kinematic phase space of the Ashtekar variables for $(\hbox{det}A)\neq{0}$, six phase space degrees of freedom per point, where the variables are diagonal.  The bonus is that all the kinematic constraints have been implemented, leaving behind the Hamiltonian constraint which in the instanton representation is easy to solve.\par  
\indent
We have shown that a nondegenerate and diagonal $A^a_i$ admits globally holonomic coordinates in the reduced theory.  Since $A^a_i$ serves also as the configuration variable for the Ashtekar phase space $\Omega_{Ash}$, it follows that on this subspace the densitized triad must also be nondegenerate.  Hence

\begin{eqnarray}
\label{THCOM}\bigl[A^f_f(x,t),\widetilde{\sigma}^g_g(y,t)\bigr]=\delta^f_g\delta^{(3)}(x,y).
\end{eqnarray}

\noindent
The conclusion is that the kinematic phase space of the dual theory must correspond the reduced phase under $(G_a,H_i)$ of the Ashtekar theory, restricted to nondegenerate triads.  Note in both phase spaces that the cotangent bundle structure has been preserved, and the two theories are equivalent when restricted to these configurations.  The bonus is that we have now implemented the initial value constraints, computed the dynamics performed a quantization, and have constructed a Hilbert space using the dual theory.

\section{Summary}

The this paper we have demonstrated the reduction of Ashtekar's theory of gravity to a kinematical phase space by implementation of the Gauss' law and the diffeomorphism constraints.  Since the initial value constraints in the reduced theory constrain only the momentum space, we were free to choose diagonal configuration space variables canonically conjugate to the densitized 
eigenvalues of $\Psi_{ae}$ in order to obtain a cotangent bundle structure.  We have deomstrated closure of the classical constraints algebra consisting of the Hamiltonian constraint, after projection to this kinematic phase space.  We have also computed the Hamiltonian dynamics on this space.\par
\indent
We then performed a quantization of the kinematic phase space, constructing a Hilbert space of states annihilated by the quantum Hamiltonian constraint for $\Lambda=0$.  These states are labelled by two eigenvalues of $\Psi_{ae}$, and appear to be consistent with the classical dynamics.  Lastly, we have clarified the relationship between of the canonical structure of the reduced theory to the original Ashtekar variables, which provides a direct route from the full Ashtekar theory to a reduced phase space for GR which can be straightforwardly quantized.  One future direction of research will be to extend the results of the present paper to 
include the $\Lambda\neq{0}$ case.

\newpage

\section{Appendix A: Expansion of the determinant on diagonal configurations}

It is convenient to factor out the leading order behaviour of the determinant of the connection from the Ashtekar magnetic field as

\begin{eqnarray}
\label{ITISEASY}
(\hbox{det}B)=(U\hbox{det}A)^2,
\end{eqnarray}

\noindent
where $U$ will be determined.  The Ashtekar magnetic field is given by

\begin{eqnarray}
\label{ASMAG}
B^i_a=\epsilon^{ijk}\partial_jA^a_k+{1 \over 2}\epsilon^{ijk}f_{abc}A^b_jA^c_k\equiv{f}^i_a+(\hbox{det}A)(A^{-1})^i_a.
\end{eqnarray}

\noindent
In (\ref{ASMAG}), $f^i_a=\epsilon^{ijk}\partial_jA^a_k$ refers to the `abelian' part and the second term is a correction due to nonabeliantiy.  We have used the fact that the $SU(2)_{-}$ structure constants $f_{abc}=\epsilon_{abc}$ are numerically the same as the Cartesian epsilon symbol in order to write the determinant, which also assumes that $A^a_i$ is nondegenerate.  Putting (\ref{ASMAG}) into the expansion of the determinant, we have

\begin{eqnarray}
\label{DETERM}
\hbox{det}B={1 \over 6}\epsilon_{ijk}\epsilon^{abc}\bigl(f^i_a+(\hbox{det}A)(A^{-1})^i_a\bigr)\bigl(f^j_b+(\hbox{det}A)(A^{-1})^j_b\bigr)\bigl(f^k_c+(\hbox{det}A)(A^{-1})^k_c\bigr)\nonumber\\
=\hbox{det}f+(\hbox{det}A)^2+{1 \over 2}\epsilon_{ijk}\epsilon^{abc}\bigl[f^i_af^j_b(A^{-1})^k_c(\hbox{det}A)+f^i_aA^a_i(\hbox{det}A)^{-1}\bigr]
.
\end{eqnarray}

\noindent
On diagonal connections the second term in (\ref{DETERM}) in square brackets vanishes, since

\begin{eqnarray}
\label{DETERM1}
A^a_if^i_a=\epsilon^{ijk}A^a_i\partial_jA^a_k=\epsilon^{ijk}(\delta^a_ia_i)\partial_j(\delta^a_ka_k)=\epsilon^{aja}a_a\partial_ja_a=0
\end{eqnarray}

\noindent
on account of the antisymmetry of the epsilon symbol.  We must now expand the first term in square brackets, evaluated on diagonal connections.  Hence we have

\begin{eqnarray}
\label{SECON}
{1 \over 2}\epsilon_{ijk}\epsilon^{abc}f^i_af^j_b(A^{-1})^k_c(\hbox{det}A)
={1 \over 4}\epsilon_{ijk}\epsilon^{klm}\epsilon^{abc}\epsilon_{cde}f^i_af^j_bA^d_lA^e_m\nonumber\\
={1 \over 4}\bigl(\delta^l_i\delta^m_j-\delta^l_j\delta^m_i\bigr)\bigl(\delta^a_d\delta^b_e-\delta^a_e\delta^b_d\bigr)f^i_af^j_bA^d_lA^e_m\nonumber\\
={1 \over 4}\bigl(f^l_af^m_b-f^m_af^l_b\bigr)\bigl(A^a_lA^b_m-A^a_mA^b_l\bigr)\nonumber\\
={1 \over 2}\bigl((f^l_aA^a_l)^2-f^m_aA^a_lf^l_bA^b_m\bigr).
\end{eqnarray}

\noindent
The first term on the right hand side of (\ref{SECON}) vanishes on diagonal connections as proven in (\ref{DETERM1}).  The second term is given by

\begin{eqnarray}
\label{SECON1}
f^l_aA^a_mf^m_bA^b_l=\epsilon^{lij}\partial_i(\delta_{aj}a_a)(\delta^a_ma_a)\epsilon^{mi^{\prime}j^{\prime}}\partial_{i^{\prime}}(\delta_{bj^{\prime}}a_b)(\delta^b_la_b)\nonumber\\
=\epsilon^{bia}\epsilon^{ai^{\prime}b}a_b(\partial_ia_a)a_a\partial_{i^{\prime}}a_b
=-{1 \over 4}\epsilon^{iab}\epsilon^{jab}(\partial_ia_a^2)(\partial_ja_b^2)
\end{eqnarray}

\noindent
where we have relabelled indices $i^{\prime}\rightarrow{j}$ on the last term.  The only nontrivial contribution to (\ref{SECON1}) occurs for $i=j$, which yields

\begin{eqnarray}
\label{SECON2}
r=-{1 \over 8}\sum_{i=1}^3I_{iab}(\partial_ia_a^2)(\partial_ia_b^2).
\end{eqnarray}

\noindent
The determinant of the Ashtekar magnetic field for a diagonal connection, which constitutes the kinematic configuration space, is given by

\begin{eqnarray}
\label{DETERM}
(\hbox{det}B)=(A^1_1A^2_2A^3_3)^2+(\partial_2A^3_3)(\partial_3A^1_1)(\partial_2A^2_2)-(\partial_3A^2_2)(\partial_1A^3_3)(\partial_2A^1_1)\nonumber\\
+(A^2_2A^3_3)(\partial_1A^2_2)(\partial_1A^3_3)+(A^3_3A^1_1)(\partial_2A^3_3)(\partial_2A^1_1)+(A^1_1A^2_2)(\partial_3A^1_1)(\partial_3A^2_2)\nonumber\\
=a_0^6e^{2T}\biggl[1+a_0^{-3}e^{-T}\bigl((\partial_2X^3)(\partial_3X^1)(\partial_1X^2)-(\partial_3X^2)(\partial_1X^3)(\partial_2X^1)\bigr)\nonumber\\
+a_0^{-2}\Bigl(e^{-2X^1}(\partial_1X^2)(\partial_1X^3)+e^{-2X^2}(\partial_2X^3)(\partial_2X^1)+e^{-2X^3}(\partial_3X^1)(\partial_3X^2)\Bigr)\biggr]\nonumber\\
\equiv{a}_0^6e^{2T}U^2,
\end{eqnarray}

\noindent
where we have defined $T=X^1+X^2+X^3$.  The end result in the full theory is that

\begin{eqnarray}
\label{SECON3}
\hbox{det}B=(\hbox{det}a)^2+r[\partial{a}],
\end{eqnarray}

\noindent
where we have defined

\begin{eqnarray}
\label{SECON4}
r=(\hbox{det}f)^2-{1 \over 8}\sum_{i=1}^3I_{iab}(\partial_ia_a^2)(\partial_ia_b^2).
\end{eqnarray}

\noindent
This fixes the definition of $U$ as

\begin{eqnarray}
\label{THISFIXES}
U=\sqrt{1+r(\hbox{det}A)^{-2}}.
\end{eqnarray}

\end{document}